\def\12{{1\over2}}
\def\bi{\bigskip}
\def\noi{\noindent}
\def\be{\begin{equation}}
\def\en{\end{equation}}
\def\bq{\begin{eqnarray}}
\def\eq{\end{eqnarray}}
\def\bc{\begin{center}}
\def\ec{\end{center}}
\def\beit{\begin{itemize}}
\def\eit{\end{itemize}}

\documentstyle[prl,aps,epsf]{revtex}
\begin{document}
\draft

\vskip 4in

\title{\Large \bf MASSES AND WIDTHS OF THE $\rho^{\pm,0}(770)$
\vskip3ex}

\author{M. Feuillat$^1$, J.L. Lucio M.$^2$\footnote{email: lucio@ifug.ugto.mx} 
\footnote{On sabatical leave from University of Guanajuato,
Mexico} and J. Pestieau$^1$\footnote{email: pestiau@fyma.ucl.ac.be}
\vspace{.5cm}} 
\address{$^1$Institute de Physique Theorique, Universite Catholique de 
Louvain \\
Chemin du Cyclotron 2, B-1348, Louvain La Neuve, Belgium. \\ 
\vspace{.5cm}
$^2$INFN Laboratori Nazionali di Frascati\\ P.O. Box 13,
I-00044 Frascati, Italy.}

\maketitle

\vskip 3.5in
\bc
{\bf \Large ABSTRACT}
\ec

\bi

\begin{abstract}
\noi 

Isospin violation in the $\rho(770)$ mass and width is considered within
the $S$ matrix approach using combined fits to the $e^+e^- \to
\pi^+\pi^-$ and $\tau^- \to \nu_{\tau}\pi^-\pi^0$ data 
performed by the ALEPH collaboration. We show that the pole
position following from the parameters obtained from the ALEPH fits 
are not sensitive to the details of the parametrization. In this context, 
we have found that the pole mass difference and the pole width difference 
between the charged and neutral
$\rho$ are consistent with zero.  We show that a one loop calculation 
including vector, axial vector and pseudo-scalar mesons can satisfactorily 
describe the observed isospin breaking. We also give an estimate for the
mass difference between the neutral and charged states of the $a_1(1260)$.

\end{abstract}

\bi\bi


\newpage

\section{Introduction}

\bi

Isospin is in very good approximation a symmetry of the strong interactions.
At the fundamental level, isospin breaking occurs as a consequence of
the electromagnetic interactions and the mass difference of the $u$ and $d$ 
quarks. The numerical value of the fine structure constant and the fact 
that $|m_u~-~m_d|$ is small compared to $\Lambda_{QCD}$ explains why this is 
a symmetry good at the few percent level. In this work we are concerned 
with the isospin and its breaking in the $\rho$ resonances. The experimental
tests of this symmetry requires the precise determination of the mass 
and width of the charged and neutral states, a non trivial task since a
precise definition of these quantities must be adopted.

The values compilated by the Particle Data Group \cite{pdg} for the 
$\rho(770)$ mass and width are spread over a wide range. This is an
unpleasent characteristic since one expects these intrinsic properties to be
independent of the process where the resonance is observed and also to be
model and background independent. This is related to the fact that
different expressions are used to fit the data leading to different 
values for the parameters -~that one is tempted to identify with the
physical mass and width~- which however are relevant only for the model
 under consideration. Although this is a practical way to deal with the 
$\rho$ meson physics,
{\it i.e.} use a given parametrization and the corresponding values for the
mass and width, the process and model independence of the physical
quantities associated to any resonance are relevant from the conceptual 
point of view. 

The analysis of the experimental data \cite{bara,ander,barkov} is 
based on theoretical assumptions such as a parametrization derived in terms of 
analyticity and unitarity \cite{sakurai}, a Breit-Wigner which incorporates
an energy dependent width reflecting the p-wave nature of the $\rho$ and 
higher isovector-vector resonances \cite{khunsa} or still
phenomenological effective field theories \cite{bena98}. There is a priori
no reason to favour any of these approaches in the description of physical 
process, on the contrary one would expect them to yield complementary or at 
least consistent information. All of them share two characteristics: {\it i)} 
they involve a momentum dependent width $\tilde \Gamma(s)$, and {\it ii)} 
they define the resonance mass $\tilde M$ as the energy for which the 
real part of the propagator is zero and the width as $\tilde
\Gamma(s=\tilde M^2)$. We will refer collectively to them as
the {\it conventional}, as opposed to the {\it pole} approach to be
discussed next. On the other hand, the $S$ matrix formalism 
\cite{eden} relates the mass and width of the resonance to pole(s) of 
the amplitude in the complex plane. The pole associated to a given resonance
is independent of the process where it is observed 
\cite{eden,sirlin}, thus this scheme offers the possibility to get an 
unambiguous determination of the intrinsic properties of the resonance. 
For the $\rho(770)$ this has been done in \cite{pesga}.

The $S$ matrix formalism not only provides a convenient
parametrization in terms of which the electromagnetic form factor of the 
pion can be expressed, but it can also be applied to any of results of the 
conventional approach. In fact we have two possibilities: the first is to 
make an expansion around the pole position  of the expressions used in the 
conventional approach and then  fit the data. We already know the result of 
such a procedure since in all cases the form factor will reduce to the 
$S$ matrix formalism, and the fit  using this scheme was performed in 
\cite{pesga}. Alternatively, fits to the data can be carried using the 
conventional parametrization and then,  given the expression used and the 
parameters obtained in fitting the data,  one can determine the associated 
pole. It is our purpose in this letter to perform the latter analysis 
considering fits reported for $e^+e^- \to \pi^+\pi^-$ and  
for the tau decay $\tau^- \to \nu_{\tau}\pi^-\pi^0$ which 
allow us to make a statement about the size of isospin violation in 
the $\rho^\pm$- 
$\rho^0$ system within a framework in which the masses and widths are
unambiguously defined.

\bi

We also consider the problem from the theoretical point
of view. To this end we take into account both possible sources of 
isospin violation: the $m_u~-m_d$ mass difference and the electromagnetic 
interactions, which is implemented at the hadron level by the use of vector 
meson dominance. The former contribution is known to vanish to lowest order
implying thus a finite result for the latter. The explicit calculation 
of the one loop electromagnetic self energy shows that a finite
result can be achieved only when the anomalous magnetic moment $\kappa$
entering the three $\rho$ vector meson vertex 
(Vector Meson Dominated $\rho-\rho-\gamma$ vertex) takes the value $\kappa=1$.
We briefly comment on this point.
\bi

\section{M AND $\Gamma$ IN THE CONVENTIONAL AND POLE APPROACHES.}
\bi

Within the 
{\bf conventional approach}, near the $\rho$ resonance, the electromagnetic 
form factor of the pion is parametrized in terms of a Breit-Wigner with a 
momentum dependent width:

\be
F_{\pi}(s) ={A(s)~\over D(s)}~;~~~~D(s)=s-\tilde M_{\rho}^2(s)+i\tilde
M_{\rho}(s)\tilde\Gamma_{\rho}(s)  \label{conven}
\en

\noi Although some expressions ocurring in the literature seem to
differ from Eq.(\ref{conven}) they can be cast in that form by an appropiated
redefinition of $A(s)$ and $\tilde\Gamma(s)$. For example if
instead of $A(s)$, we use $ \bar A(s)=A(s)(1-ix(s))$,  $D(s)$ should be
changed to:

\bc 
\be

\bar D(s)=s-\tilde M_{\rho}^2(s)+\tilde M_{\rho}(s) \Gamma_{\rho}(s) x(s)+
i(\tilde M_{\rho}(s)\tilde \Gamma_{\rho}(s)-(s-\tilde M_{\rho}^2(s))x(s)).  
\label{scale}
\en 
\ec

\noi Since in the conventional approach  the mass and width of the
resonance are defined respectively by

\be
Re( D(m_{\rho}^2))=0~;~~~~and~~~~Im(D(m_{\rho}^2)=m_{\rho}\Gamma_{\rho},
\en

\noi the freedom in the choice of parametrization (compare
Eqs.(\ref{conven},\ref{scale})) implies  an arbitrariness 
in the mass and width deffinition. On the other hand, the {\bf $S$
matrix approach} \cite{eden,sirlin} avoids this problem  by considering a 
simple  observable pole with $s$ independent width and residue:
\be
F_{\pi}(s)={R_p~\over s-s_p}+B(s),
\en

\noi where $s_p$ is the pole position, $R_p$ the pole residue and 
$B(s)$ is the remaining part around the pole. In order that this description makes
sense the remaining part $B(s)$ (which is fixed by the fit to the data) should be
a soft function of $s$ affecting minimally the pole position, except for
obvious resonance effects as in $\rho-\omega$ mixing \cite{pesga}. In this 
formalism the mass $M$ and width $\Gamma$ of the resonance are deffined 
through:

\be
s_p=M^2-iM\Gamma.        \label{pole}
\en

\noi The fact that the background is a soft function arround the
pole (in this case for $\sqrt{s}~\approx~m_{\rho}^2$), does not imply that it
can not have poles at other energies,  however these can be included in an 
appropiated way in the background. Different parametrizations within the 
conventional approach
correspond to different backgrounds in the $S$ matrix formalism. Thus
even though the mass and width values obtained from a given ``conventional'' 
treatment are only relevant for that particular parametrization, on physical
grounds one expects that all of them have a pole in the same position 
providing thus, through Eq.(\ref{pole}), the physical mass and width of the 
resonance. Notice that in finding the pole $D(s_p)=0$ no information about 
the overall normalization nor about $A(s)$ is required, as one would expect
from the $S$ matrix approach.

\bi
The cleanest determination of the $\rho$ mass and width comes from the
$e^+e^-$ anihilation and tau-lepton decay data, which are in agreement 
with each other up to an overall normalization factor \cite{pdg}. 
Although several analysis for the  $e^+e^- \to \pi^+\pi^-$
\cite{bara,barkov,bena98,gard,bena93} and $\tau^- \to\nu_{\tau}\pi^-\pi^0$ 
\cite{bara,ander} data have been performed, we  concentrate in the
results of the  ALEPH collaboration \cite{bara} which seems to 
us the best suited for a study of isospin breaking since a combined fit to 
both sets of data was carried by these authors. Notice however that the 
agreement is excellent with the fit to the data of the high statistics 
experiment by the CLEO collaboration \cite{ander}. The $\rho^\pm$,
$\rho^0$ masses and 
widths resulting from the analysis in \cite{bara}  and the values
associated to the corresponding pole position are summarized in Table 1
 for the masses and Table 2 for the widths. 

\bi

\renewcommand{\arraystretch}{1.3}
\bc
\begin{tabular}{|c|c|c|c|c|c|c|} \hline
Fit& \multicolumn{2}{c|}{$m_{\rho^\pm}(MeV)$}
&\multicolumn{2}{c|}{$m_{\rho^0}(MeV)$}&\multicolumn{2}{c|}
{$m_{\rho^\pm}-m_{\rho^0}(MeV)$} \\
\cline{2-7}
~~&conven.&~~~~~~~pole~~~~~~~&conven.&~~~~~~~pole~~~~~~~&~~~conven~~~&
~~~~~pole~~~~~ \\
\hline\hline
KS($\lambda=1$)&$~~~773.4 \pm 0.9~~~
$&$757.0 \pm 1.3$& ~~~$773.4 \pm 0.7~~~$&$756.9 \pm 1.0$ &
$0.0 \pm 1.0$&$0.1 \pm 1.6 $ \\
GS($\lambda=1$)&$775.7 \pm 0.9$&$758.1 \pm 1.3$&$775.7 \pm 0.7$&$758.0 \pm
1.0$&$0.0 \pm 1.0$&$0.1 \pm 1.6 $ \\
GS($\lambda=0.45 \pm 0.11$)&$783.8 \pm 3.0$&$758.3 \pm 5.4$&$783.8 \pm
3.0$&$758.1 \pm
5.4$&$0.0 \pm 1.2 $&$0.2 \pm 7.6$ \\
\hline
\end{tabular}
\ec
\hspace{0.65 cm}
\begin{minipage}{16.3 cm}
{\small Table 1. Masses of the $\rho^\pm$ and  $\rho^0$ resonances as 
obtained by  Barate {\it et al} \cite{bara} (second and fourth column) 
from a combined 
fit to  $e^+e^- \to \pi^+\pi^-$ and $\tau^- \to \nu_{\tau}\pi^-\pi^0$
data. For comparison in the third and fifth column 
we quote the values associated to the corresponding pole.
Details about the notation can be found in the quoted literature.} 
\end{minipage}

\renewcommand{\arraystretch}{1.3}
\bc
\begin{tabular}{|c|c|c|c|c|c|c|} \hline
Fit& \multicolumn{2}{c|}{$\Gamma_{\rho^\pm}(MeV)$}
&\multicolumn{2}{c|}{$\Gamma_{\rho^0}(MeV)$}&\multicolumn{2}{c|}
{$\Gamma_{\rho^\pm}-\Gamma_{\rho^0}(MeV)$} \\
\cline{2-7}
~~&conven.&~~~~~~~pole~~~~~~~&conven.&~~~~~~~pole~~~~~~~&~~~conven~~~&
~~~~~pole~~~~~  \\
\hline\hline
KS($\lambda=1$)&$~~~147.7 \pm 1.6~~~
$&$143.2 \pm 1.5$& ~~~$147.3 \pm 1.3~~~$&$142.7 \pm 1.2$ & $0.4 \pm 1.0$ &
$0.5 \pm 1.9 $ \\
GS($\lambda=1$)&$150.8 \pm 1.7$&$145.3 \pm 1.5$&$150.8 \pm 1.3$&$145.2 \pm
1.2$&$0.0 \pm 2.0$  &$0.1 \pm 1.9 $ \\
GS($\lambda=0.45 \pm 0.11 $)&$162.0 \pm 5.3$&$145.1 \pm 6.3$&$162.4 \pm
5.0$&$145.2 \pm
6.5$&$-0.4 \pm 2.5$&$-0.1 \pm 8.7$ \\
\hline
\end{tabular}
\ec
\hspace{0.65 cm}
\begin{minipage}{16.3 cm}
{\small Table 2. Widths of the $\rho^\pm$ and  $\rho^0$ resonances as 
obtained by  Barate {\it et al} \cite{bara} (second and fourth column) 
from a combined 
fit to  $e^+e^- \to \pi^+\pi^-$ and $\tau^- \to \nu_{\tau}\pi^-\pi^0$ data.
For comparison in the third and fifth column we quote the values associated 
to the corresponding pole. Details about the notation can be found in the 
quoted literature.} 
\end{minipage}

\bi
\bi

The results are remarkable. While in the conventional approach the masses
are spread over a 10 MeV range, the $S$ matrix formalism leads values
that differ at most in 1.3 MeV. Similarly for the width where the two 
approaches yield values differing, within each scheme, in 14 and 2 MeVs 
respectively. We can safely conclude that in contradistinction to the
conventional approach the pole position is independent 
of the details of the parametrization used to fit the data providing thus
an unambiguous definition for the mass and width of the $\rho$.
It is worth remarking that the mass and width values we obtained applying
the $S$ matrix formalism to different parametrizations of the conventional
approach agree with previous anlysis performed fully in the $S$ matrix
formalism \cite{pesga,bernicha,bena99,tesis}
but are systematically below the values obtained in the 
conventional approach and the value reported by the PDG \cite{pdg}.

\bi

As far as the isospin breaking is concerned, both schemes lead similar
results. Although in the conventional approach there is a strong dependence
upon the parametrization used to fit the data, it is the same for the
charged and neutral channels and it cancels out when differences are
considered. Let us emphasize again that the advantage of the pole approach is
that it is based on an unambiguous definition of the mass and width and so,
besides a nearly vanishing $\rho^{\pm}-\rho^0$ mass difference we have a 
precise knowledge of each mass and width. The isospin violation in the
coupling constants, $g_{\rho}$, implied by the pole decay rates is also 
consistent with zero, $g^2_{\rho^{\pm}}$ and $g^2_{\rho^0}$ are equal up to 
one percent.

\bi

\noi When applying the $S$ matrix formalism to one of the parametrizations
used in the conventional approach, as we have done, the uncertainty in the
pole position follow from the result of the
original fit. In order to calculate the errors quoted in
Tables 1 and 2, we used the following procedure: we obtain the pole
position using the central value of the parameters obtained from the fit,
we then determine the variation of the pole position when the parameters of
the fit are changed by one standard deviation. We quote for the pole
position error the maximun variation found in this way. One further comment
regards the $\lambda$ parameter \cite{bena93} used in the second version of the
Gounaris-Sakurai parametrization \cite{bara}. $\lambda$ modify the $s$
dependence of the $\rho$ decay width and it also  changes the analytical 
structure obtained in an effective field theory. Notice that the errors 
are much larger when $\lambda$ is left as a free parameter, {\it i.e.} when
the $s$ dependence is not the one predicted by the effective field theory \cite{bena98} .

\bi

\section{$m_{\rho^\pm}-m_{\rho^0}$ MASS DIFFERENCE WITHIN A VECTOR MESON 
DOMINANCE MODEL.}
\bi

In this section we consider the theoretical evaluation of the $\rho^
\pm~-~\rho^0$ mass difference. In order to calculate the contribution due
to the $m_u-m_d$ mass difference, we need to evaluate the matrix element of 
the flavour symmetry breaking Hamiltonian $H_B~=~m_u \bar uu~+~m_d \bar dd~+~m_s \bar
ss$. The matrix elements between $\rho$ states can be written as:
\bc
\bq
&<\rho|&(m_u \bar uu~+~m_d \bar dd~+~m_s \bar ss)|\rho>~=~{1 \over
  \sqrt{2}}(m_u-m_d)<\rho|{\bar uu~-~\bar dd \over \sqrt{2}}|\rho>
\nonumber  \\ 
+{1 \over \sqrt{6}}(m_u+&m_d&-2m_s)<\rho|{\bar uu+ \bar dd-2 \bar ss \over
  \sqrt{6}}|\rho> 
+{1 \over \sqrt{3}}(m_u+m_d+m_s)<\rho|{\bar uu+ \bar dd+ \bar ss \over
  \sqrt{3}}|\rho>.
\eq 
\ec

\noi The operators entering in the two last matrix elements are singlet under 
$SU(2)$, therefore they give the same contribution to the $\rho^ \pm$
and $\rho^0$ and thus, do not contribute to the mass difference. On the
other hand the first matrix element vanishes identical as consequence of
$G$ parity. We therefore conclude that to lowest order in ($m_u-m_d$) the 
$\rho^ \pm-\rho^0$ mas difference vanishes:

\bc
\be
m_{\rho^\pm}~-~m_{\rho^0}  |_{ m_u \neq m_d}~=~0.
\en 
\ec

\bi

This result has an important implication. Since the $m_{u,d}$ quark masses
have completely disappeared from our calculations, there are no fundamental
constants of the underlying QCD theory which could be used to renormalize,
{\it i.e.} to absorbe possible divergencies appearing in the
calculation. Therefore the electromagnetic self-mass contribution must be
finite \cite{thoma}. Later on we will see that the same conclusion can
and has been reached \cite{harari} using different arguments.

\bi

Let us now consider the second source of isospin breaking, the 
electromagnetic interactions. The lowest order contributions are shown
in Fig. 1($a-e$), where $P$ stands for a pseudoscalar meson which in our
case can be either $\pi$, $\eta$ or $\eta\prime$, $a_1$ denotes an axial
vector and
$V=\rho, \omega$ or $\phi$ enters through Vector Meson Dominance.
The first diagramm is non vanishing for the charged 
$\rho$ and vanishes for the $\rho^0$. Diagram $b$ is non vanishing 
($V=\omega$ or $\phi$) both for the charged and neutral contribution.
Notice furthermore that the $\eta$ or $\eta\prime$ can be
intermediate state only for the neutral $\rho^0$ self-energy. We will
neglect the contribution of the $\eta$ and $\eta\prime$ as we expect these 
to be small (one hundred times smaller than the contribution
of diagram $e$). On the other
hand, the contribution from the loop with an internal pion yields the same
contribution for the charged and neutral $\rho$'s. The third diagramm
vanishes identically when $V=\rho$ due to $G$ parity and gives the same
contribution to the $\rho^\pm$ and $\rho^0$ self-mass when $V=\omega$ or $\phi$ and
therefore does not contribute to the mass difference. Figures $d$ and $e$
are non-vanishing and contribute only to the charged and neutral meson mass
respectively. Thus, in the model we are
considering, the $\rho^\pm$~-~$\rho^0$ mass difference is given by the
contributions shown in Fig. 1 ($a,d$ and $e$).
To describe the photon $\rho$ vertex we use Vector Meson
Dominance, so that the $\rho$ form factor is modeled by a $\rho$
propagator. The Feynman rules for the electromagnetic interaction of a
charged massive spin $S=1$ vector meson can be found in \cite{yang}.
These rules depend upon the anomalous magnetic moment $\kappa$ of the 
vector meson. Using these rules, we find that the correction to the 
$\rho^\pm$ mass is given by (for later use we introduce obvious notation
for the loop integrals of $T_1, T_2$ and $T_3k_{\alpha}k_{\beta}$): 

\vspace{2.0cm}

\centerline{\epsfxsize = 350 pt
\epsfbox{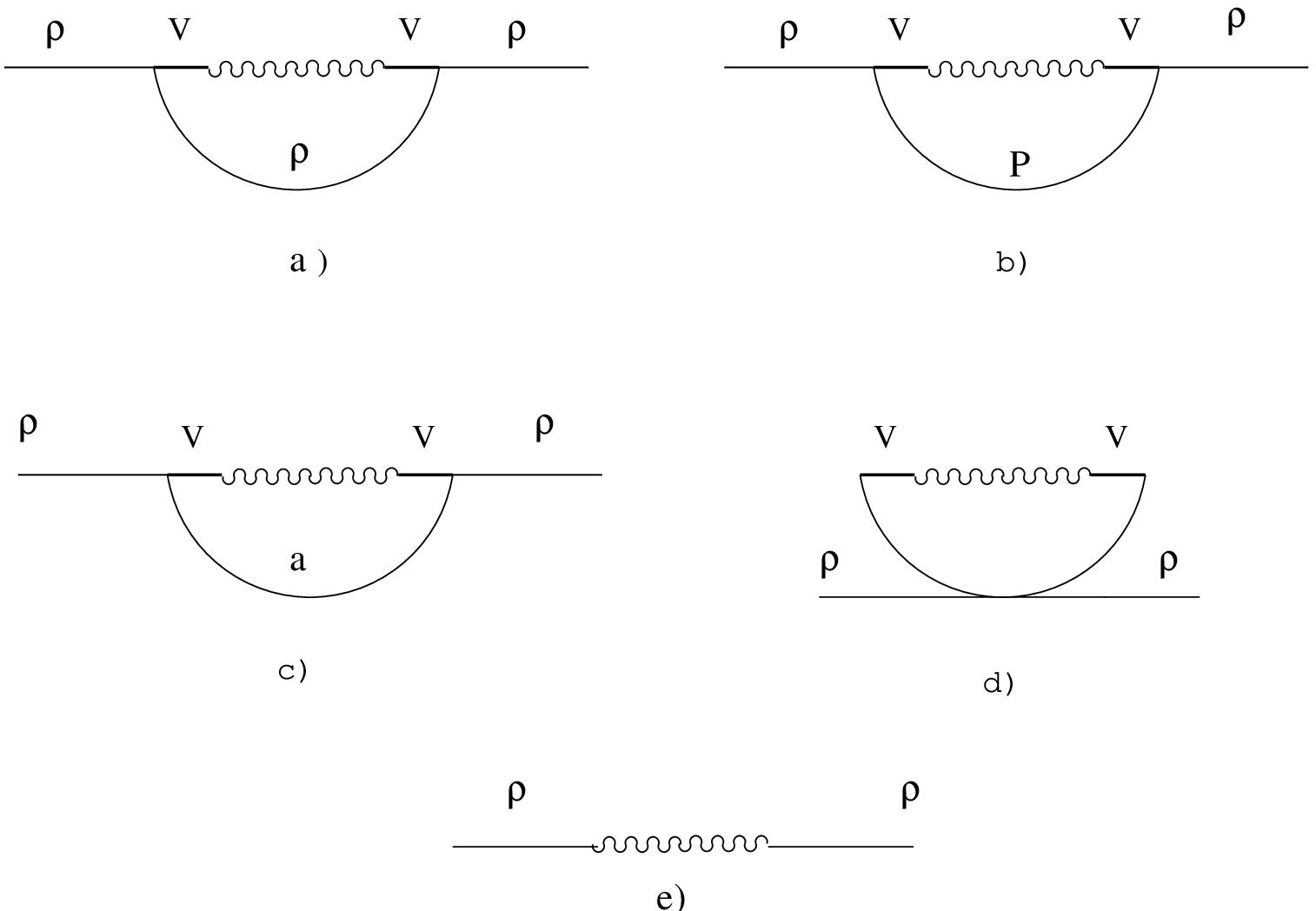}}
\begin{center}
{\small {Fig. 1}  \\

\bi

\noi One loop electromagnetic contribution to the self-mass of the $\rho$. 
Vector Meson Dominance is used to describe the coupling of the photon to the
hadrons. P stands for a pseudoscalar ($\pi$,  $\eta$ or $\eta\prime$), V for a vector
meson ($\rho$, $\omega$ or $\phi$) and $a$ for the axial vector $a_1(1260)$. See main
text for further details.}

\end{center}

\bi
\bi

\bc
\bq
i\delta^{\gamma}_{m^2_{\rho^\pm }}~&=& ~-e^2 M^4 \epsilon^{\alpha}_i(p)
\epsilon^{* \beta}_i(p)\int\!  {d^4k \over (2 \pi)^4}\left(T_1~g_{\alpha\beta}
~+~T_2~g_{\alpha\beta}~+~T_3~k_{\alpha}k_{\beta} \right) \\

& =&{\alpha M^4  \over 4 \pi}(t_1+t_2+t_3) 
\nonumber
\eq  
\ec
where:

\bc
\bq
T_1~=~{4m^2 \over
    k^2\left[k^2-M^2\right]^2\left[k^2+2pk\right]}-{2 \over
      k^2\left[k^2-M^2\right]^2}-{1 \over \left[k^2-M^2\right]^2 \left[
k^2+2pk \right]}, 
\eq 
\ec

\bc 
\bq
\nonumber
T_2~=~{\kappa^2 \over
    \left[k^2-M^2\right]^2\left[k^2+2pk\right]}-{\kappa^2k^2 \over
      4m^2\left[k^2-M^2\right]^2\left[k^2+2pk \right]}-{3\kappa^2-8\kappa+4
 \over 4m^2\left[k^2-M^2\right]^2}, 
\eq
\ec

\bc
\bq
\nonumber
T_3~=~{\kappa^2+4\kappa-4 \over
    k^2\left[k^2-M^2\right]^2\left[k^2+2pk\right]}+{\kappa^2-2\kappa+1 \over
      m^2\left[k^2-M^2\right]^2 \left[
k^2+2pk \right]}.

\eq
\ec

\noi Remark that the $T_1$ contribution to the integral is exactly the same
entering in the calculation of the electromagnetic mass difference of the
pion \cite{thoma}. Notice furthermore that as it stands, the expression
involves ultraviolet divergences. For completness we quote the analytical
expression for $\delta_{m_{\rho}^2}$ as a function of $\kappa$ and the
$\rho^{\pm} (m)$ and $\rho^0 (M)$ masses:

\bi

\bc \bq
t_1~=~{2 \over M^2}~+~
{1 \over 2m^2}ln{M^2 \over m^2}~+~{M^2~+2m^2 \over m^2M^4}R~tan^{-1}({R 
\over M^2}),
\eq \ec

\bc
\bq
\nonumber
t_2~=~-{\Delta \over
  m^2}(\kappa^2-2\kappa+1)-{\kappa^2 \over 4m^2}\left[2+(3-{M^2 \over
  m^2})ln{M^2 \over m^2}+ 
 {2(M^2-4m^2)(M^2-m^2) \over m^2R}~tan^{-1}({R \over M^2})\right],
\eq \ec

\bc
\bq
\nonumber
t_3~&=&~{(\kappa^2-2\kappa+1) \over 4m^2}\left[3+\Delta +{M^4-4m^2M^2+2m^4 
\over 2m^4}ln{M^2 \over m^2}-{M^2 \over m^2}-
{2m^2-M^2 \over M^4}R~tan^{-1}({R \over M^2})\right] \\

\nonumber
&+& {(\kappa^2+4\kappa-4) \over 4m^2} \left[
-{2 \over 3}+ {M^2-3m^2 \over 3m^2}ln{M^2 
\over m^2}-{2(m^2-M^2)(4m^2-M^2) \over 3M^2R}~tan^{-1}({R \over M^2}) \right].
\eq
\ec

\bi
with:
\bi
\bc 
\bq

\Delta={2 \over 4-n} -\gamma +ln(4\pi)-ln{M^2 \over \mu^2}
~~~~~~~~~~~~R=\sqrt{M^2(4m^2-M^2)}.
\eq 
\ec

\noi  The self-energy involves ultraviolet divergencies -
contained in the $\Delta$ factors - which are absent for $\kappa=1$. 
On the other hand the finiteness of the $\rho$ electromagnetic self-mass 
is required not only from the reasoning presented paragraphs above. 
In Ref. \cite{harari} the
same conclusion was reached using completely different arguments (based on
dispersion relations and Regge pole theory). Furthermore one expects the 
$\rho-\rho-\rho$ vertex to be
symmetric in the indices, which is achieved only if $\kappa~=~1$. Taking
that value for $\kappa$ and $m~=~M~=~757.5 ~ MeV$ we get:

\bi

\bc 
\be
\delta^\gamma_{m_{\rho^ \pm}}~=~{\delta^\gamma_{m^2_{\rho^ \pm}} \over
  2m_{\rho}}~=~{\alpha m_\rho \over 8\pi}\left\{2
    +\pi\sqrt{3}- {2 \over 3} \right\}~=~1.49~ MeV.   \label{numero}
\en 
\ec

\bi

On the other hand, in the model we are considering the correction to the 
$\rho^0$ meson mass is given by the diagramm in Fig.1($e$).  which leads to:

\bi

\bc
\be
\delta^{\gamma}_{m_{\rho}}~=~{\delta^{\gamma}_{m^2_{\rho^0}} \over 
2m_{\rho^0}}
~=~{2\,\pi\,\alpha \, m_{\rho^0} \over f^2_{\rho}}
\en  
\ec

\noi With $4.83~<~f_{\rho}~<~4.91$ (as determined in Ref.(\cite{bernicha}))
from fits to $e^+e^- \to \pi^+\pi^-$ data in the $\rho$ region in  the 
framework of the $S$ matrix formalism, we use the central value  
$f_{\rho}=4.87$ so that putting all togheter we finally obtain:

\bi

\be
\Delta m_{\rho}~=~m_{\rho^\pm}~-~m_{\rho^0}~=~\delta_{m_{\rho^\pm}}-
\delta_{m_{\rho^0}}=~0.02 \pm 0.02. 
\en
\bi

\noi Our result is consistent with a previous calculations by  Bijnens and 
Gosdzinsky \cite{bijnens} who using a different approach find $-0.7 MeV~ 
<~ m_{\rho^{\pm}}- m_{\rho^0}~ <~ 0.4 MeV$.

\bi

A similar analysis can be carried for the $a_1(1260)$.To obtain an estimate
we neglect the analog of diagram in Fig. $1b$ which contributes to the self 
mass of the charged but not of the neutral $a_1(1260)$. 
$G$ parity requires $V=\rho$ and the direct  coupling of the photon 
to the neutral $a_1$ (Fig. $1e$) is not alllowed.
The contribution to the $a_1(1260)$ mass difference from diagrams $1a$ and
$1d$ is obtained replacing $m_{\rho}$ by $m_a$ in Eq. (\ref{numero}), which
leads to:

\bc
\be 
\Delta m_a=\delta m_{a^{\pm}} -  \delta m_{a^0}= \delta m_{a^{\pm}}=2.42
MeV.
\en
\ec

\bi

\section{SUMMARY}
\bi

In this work we considered the difference in masses and widths of the charged
and neutral $\rho(770)$ mesons. The $S$ matrix formalism provides a
framework where these intrinsic properties are unambiguously defined.
The determination of the mass and width of the $\rho$ in that framework
have been presented in Ref. \cite{pesga,bernicha}. Here we go one step
further by taking as starting point the combined fit to the 
 $e^+e^- \to \pi^+\pi^-$ and $\tau^- \to \nu_{\tau}\pi^-\pi^0$ data
performed by the ALEPH collaboration \cite{bara}. The anlaysis in
\cite{bara} was carried using the Khun-Santamaria \cite{khunsa} formalism
and two versions of the Gounaris-Sakurai \cite{sakurai} parametrization 
of the pion
form factor. We have shown that the pole position following from the 
parameters obtained from the fit to the data are not sensitive to the
details of the parametrization. In this context, we have found that the 
mass difference between the charged and neutral $\rho$ is consistent with 
zero very much as the width is. Our results for the $\rho$ mass and width 
agree  with the pole position found when the pion form factor is
parametrized from the very begining according to the $S$ matrix formalism 
and are smaller than the values reported by the PDG.

\bi

{From} the theoretical point of view, we considered the isospin violation
induced by the $m_u-m_d$ mass difference which is known to vanish to
the lowest order. We have argued that the electromagnetic self-mass should
be finite and we have shown that this  result is
obtained when the elctromagnetic coupling of the $\rho$ is modeled using
vector meson dominance and the $\rho$ anomalous magnetic moment 
$\kappa$ takes the value $\kappa=1$. The model satisfactorily describes 
the observed isospin breaking.

\bi

Let us finally comment that  the parametrization independence of the pole
position holds when a larger set of fits is considered. Indeed when the
typical parametrizations used in Ref.\cite{bara,ander,barkov,bena98,bena93}
are analyzed along the lines of this work (except for the fit called 
VMD2(NU) \cite{bena98}, which as
pointed out by the author leads to a $\chi^2=148/77$ compared to the
$\chi^2 \approx 1$ of all the others fits)
the mass  values associated to the pole are spread over 
a five MeV region arround 758 MeV, in contrast to the 
25  MeV arround 767 MeV obtained in the conventional approach. A similar
situation holds for the width where the spreads are respectively  7 and
22 MeV and the central values 143 and 151 MeV respectively.

\section{ACKNOWLEDGMENTS}
\bi
\noi We thank M. Benayoun and G. Lopez-Castro for observations. J.L.L.M acknowledges 
finantial support from CONACyT and the hospitality of the Institute de Physique 
Theorique UCL where part of this work was done.

\bi

\bi

\end{document}